\documentclass[3p,times]{elsarticle}
\usepackage{amssymb}
\usepackage[figuresright]{rotating}

\usepackage{amsmath}
\usepackage[english]{babel}
\usepackage{graphicx}
\usepackage{tikz}
\usetikzlibrary{patterns}

\begin{document}

\begin{frontmatter}

\title{Scaling Properties of Parallelized Multicanonical Simulations}
\author[]{Johannes Zierenberg}
\ead{johannes.zierenberg@itp.uni-leipzig.de}
\author[]{Martin Marenz}
\ead{martin.marenz@itp.uni-leipzig.de}
\author[]{Wolfhard Janke}
\ead{wolfhard.janke@itp.uni-leipzig.de}
\address{Institut f\"ur Theoretische Physik,
         Universit\"at Leipzig, Postfach 100920, D--04009 Leipzig, Germany}

\begin{abstract}
  We implemented a parallel version of the multicanonical algorithm and applied it to a 
  variety of systems with phase transitions of first and second order. The parallelization 
  relies on independent equilibrium simulations that only communicate when the multicanonical
  weight function is updated. That way, the Markov chains efficiently sample the temporary 
  distributions allowing for good estimations of consecutive weight functions.

  The systems investigated range from the well known Ising and Potts spin systems
  to bead-spring polymers. We estimate the speedup with increasing number of parallel processes.
  Overall, the parallelization is shown to scale quite well.
  In the case of multicanonical simulations of the $q$-state Potts model ($q\ge6$) and
  multimagnetic simulations of the Ising model, the optimal performance is limited due to 
  emerging barriers.
\end{abstract}


\end{frontmatter}


Umbrella sampling algorithms like the multicanonical method~\cite{BergMUCA,Janke1998} and the 
Wang-Landau method~\cite{WangLandau} have been applied to a variety of complex systems over the last
two decades. They are well suited for the investigation of phase transitions, especially of 
first order and may be applied to systems with rugged free-energy landscapes in physics, 
biology, and chemistry~\cite{Reviews}.

For complex systems, a carefully chosen set of Monte Carlo update moves is usually the key
to a successful simulation. But with computer performance increasing mainly in terms of parallel
processing on multi-core architectures, it is of advantage when the algorithm can be parallelized.
This has been done recently for the Wang-Landau recursion~\cite{Zhan2008,Landau2012} and for the
standard multicanonical recursion~\cite{ParallelMUCA,Zierenberg2013}. While the former implementation
relies on shared memory, introducing racing conditions and frequent communication, the latter
benefits from independent Markov chains with occasional communication. 

After a short summary of the parallel multicanonical method, we will proceed with a demonstration
of its performance for several spin systems and a flexible polymer.

\vspace{1em}
The multicanonical (MUCA) method  can be applied to a variety of ensembles. Still, it is probably 
easiest to understand using the example of the canonical ensemble. For a fixed temperature, all
configurations that the system may assume are weighted with the Boltzmann weight $P(E)=\exp[-\beta E]$,
resulting in a temperature dependent energy distribution. The idea is to replace this Boltzmann weight
by an arbitrary weight function $W(E)$, which may be modified iteratively such that the resulting energy 
distribution covers the full range of canonical distributions, hence it is called multicanonical. 
In terms of the partition function, this may be written as
\begin{equation}
              Z_{\rm can}  = \sum_E \Omega(E)e^{-\beta E}
  \rightarrow Z_{\rm MUCA} = \sum_E \Omega(E) W(E),
\end{equation}
where $\Omega(E)$ is the density of states. The canonical distributions and expectation values
are recovered by reweighting:
\begin{equation}
 \langle O \rangle_{\beta} = \frac{\langle O_i e^{-\beta E_{i}}W^{-1}\left(E_{i}\right)\rangle_{\rm MUCA}}
                                  {\langle     e^{-\beta E_{i}}W^{-1}\left(E_{i}\right)\rangle_{\rm MUCA}}.
\end{equation}
The most difficult task is the weight modification, which requires some effort. 
The easiest way is to construct consecutive weights from the last weights and the
current energy histogram: $W^{(n+1)}(E) = W^{(n)}(E)/H^{(n)}(E)$.
More sophisticated methods use the full statistics of previous iterations
for a stable and efficient approximation of the density of states~\cite{Janke1998}.
All our simulations use the latter version implemented with logarithmic weights in order
to avoid numerical problems.

\begin{figure}
  \centering
  \includegraphics[width=9cm]{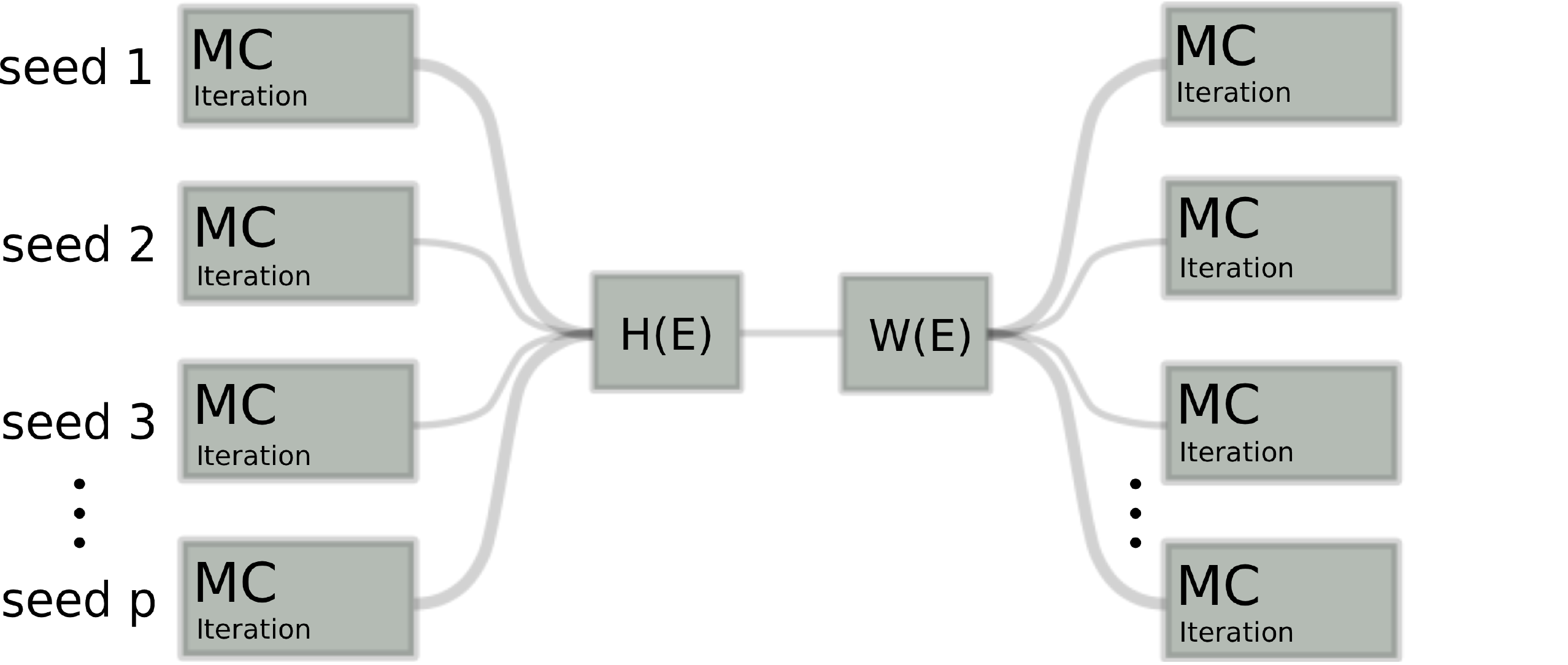}
  \caption{ Scheme of the parallel implementation of the multicanonical algorithm on $p$ cores.
            After each iteration with independent Markov chains (but identical weights), the
            histograms are merged, the new weights are estimated and distributed to all processes again.
          }
  \label{fig:pmuca}
\end{figure}

The basic idea of the parallel implementation is shown in Fig.~\ref{fig:pmuca}. The system is initialized
in $p$ independent realizations with the same weight function, which are distributed onto different cores.
After each iteration, the individual histograms are merged and provide an estimate of the distribution
$H^{(n)}$ belonging to the current weight function $W^{(n)}$. 
This is used to determine the consecutive weight function $W^{(n+1)}$, which is again distributed to all $p$ 
cores. The whole procedure is repeated until the weight function results in a flat energy distribution.
Since the weight modification is usually very fast compared to a single iteration, communication is kept
to a bare minimum.

Moreover, the parallelization may easily be applied to other ensembles, for example multimagnetic 
(MUMA) simulations. In this case, the coefficients of the partition function are modified by a
correction weight function, which depends for example on the magnetization and is again modified 
iteratively in order to yield a flat histogram in the parameter:
\begin{equation}
              Z_{\rm can}  = \sum_E \Omega(E)e^{-\beta E}
              \rightarrow Z_{\rm MUMA} = \sum_{E,M} \Omega(E,M)e^{-\beta E}W(M).
\end{equation}
The parallelization is completely analogous to the standard case and the performance of the 
parallel multimagnetic simulation will be demonstrated below.

\vspace{1em}
For a fair comparison of the performance of this parallel implementation, we need to consider
a few aspects. First of all, the parallelization relies on independent Markov processes, 
which leads to different simulations for different degrees of parallelization. Consequently, we 
may only compare the average performance per degree of parallelization. 
Furthermore, a number of parameters can influence the results and need to be fixed.
One example is the number of sweeps per iteration. In order to provide a proper estimate of the 
performance, we determined the optimal number of sweeps per iteration $M_{\rm opt}$ for all degrees
of parallelization in the cases of the Ising model and the $8$-state Potts model.
A detailed description can be found in \cite{Zierenberg2013}. The results of this analysis are 
\begin{equation}\label{eq:powerLawDependence}
 \begin{aligned}
  M_{\rm opt}^{\rm (Ising) }(L,p) &= 5.7(5) \times L^{2+0.51(4)}\frac{1}{p} \\
  M_{\rm opt}^{\rm (8Potts)}(L,p) &= 24(4)  \times L^{2+0.67(6)}\frac{1}{p}.
 \end{aligned}
\end{equation}
We extrapolated this result for the remaining spin systems.
Another factor to consider is the thermalization time. In order to remove additional parameters
in our investigation, we decided to thermalize only in the beginning and not in between iterations.

Since the parallelization changes the outcome of the simulation, also the
number of iterations until convergence is influenced. This allows us to consider two different
measures of the performance. 
One is given by the speedup in convergence time, comparing the time $t_p$ a $p$-core simulation needs
until convergence of the MUCA weights with the time $t_1$ a single-core simulation needs:
\begin{equation}
  S_p = \frac{t_1}{t_p}.
\end{equation}
Obviously, this answers the question of required simulation time with increasing parallelization
but depends strongly on the involved hardware. Thus, the result may differ if investigated on different
compute clusters.
Another possibility is to consider a time-independent statistical speedup by comparing the total number 
of sweeps per core until convergence. As the optimal number of sweeps per iteration $M_{\rm opt}(p)$ is
fixed for all realizations, this results in measuring the average number of iterations 
until convergence $\bar{N}_{\rm iter}$.
\begin{equation}
    S^*_p = \frac{[ \bar{N}_{\rm iter} M_{\rm opt}(1)]_{1}}{[ \bar{N}_{\rm iter} M_{\rm opt}(p)]_{p}},
\end{equation}
In the following, we will use the Ising model to demonstrate the differences. Afterwards we will restrict
ourselves to the time-independent statistical speedup for simplicity.

\begin{figure}
  \includegraphics{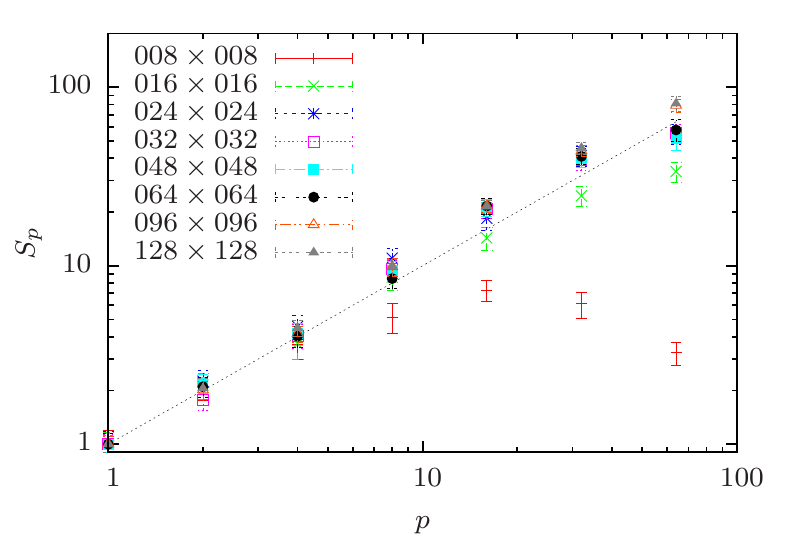}
  \includegraphics{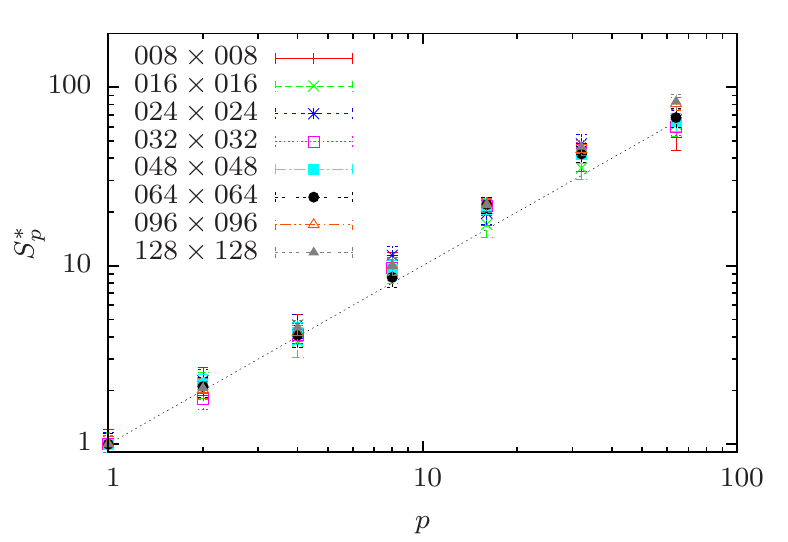}
  \caption{Performance in case of the Ising model for different system sizes: 
           (left) the speedup in real time and (right) the speedup in statistics.
          }
  \label{fig:ising}
\end{figure}

\vspace{1em}
We consider the two-dimensional Ising system as a first test case. This spin model with 
nearest-neighbor interaction exhibits a temperature driven second-order phase transition.
The Hamiltonian is defined as $\mathcal{H} = -J\sum_{\langle i,j\rangle} s_i s_j$.
Figure \ref{fig:ising} shows the performance of the method. It can be seen that the statistical
speedup scales nicely for all system sizes, in fact $S^*_p\simeq p$. This means that the total 
statistics is efficiently distributed onto all cores.
The time speedup also scales well, except for the small system sizes where the duration of a single 
iteration was of the order of milliseconds for which our network communication is insufficient.

\begin{figure}[h]
  \centering
  \includegraphics{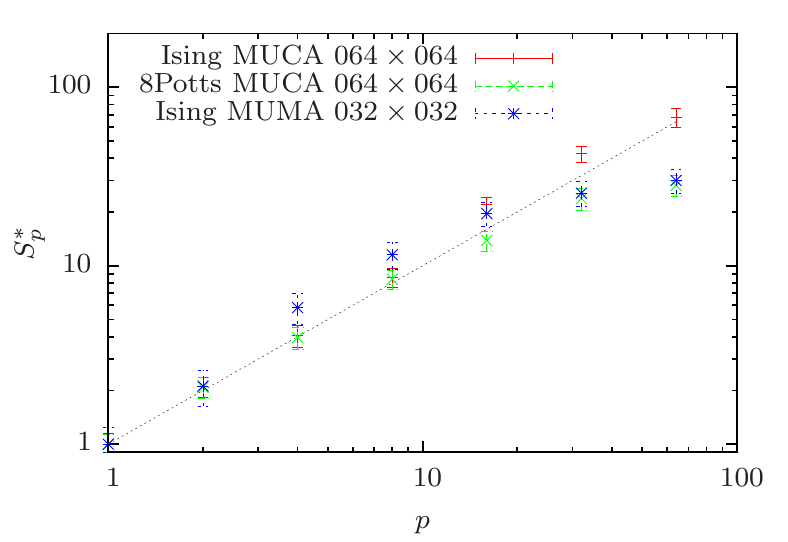}
  \caption{Statistical speedup of selected spin systems.
          }
  \label{fig:spinsystems}
\end{figure}

The two-dimensional $q$-state Potts model is described by 
$\mathcal{H} = -J\sum_{\langle i,j\rangle} \delta(s_i, s_j)$, where $s_i \in \left\{0, \dots ,q-1\right\}$
and interaction is restricted to nearest neighbors. The system shows a temperature driven 
first-order phase transition for $q\ge5$ and a second-order phase transition otherwise.
Applying the parallel multicanonical method to the \mbox{$8$-state} Potts model demonstrates its
effect on systems with first-order phase transitions. Indeed, also in this case the parallelization
works well, but with increasing degree of parallelization the speedup seems to saturate 
(see Fig.~\ref{fig:spinsystems}). 
A similar effect is observed when applying the parallelization to a multimagnetic simulation
of the Ising system, also shown in the figure. In this case, the Ising system is simulated at fixed
temperature $T=\frac{2}{3}T_C$, while it is attempted to achieve a flat distribution of the magnetization. 
The occurring field-driven phase transition is of first order.
The saturation of the speedup may be explained by large integrated autocorrelation times
accompanying concealed barriers. Thus, when reducing the number of sweeps per core with increasing 
degree of parallelization this might reach a point where the individual sweeps are too short in order
to efficiently cross emerging concealed barriers.
For a detailed description we refer to \cite{Zierenberg2013}.

\begin{figure}[h]
  \centering
  \includegraphics{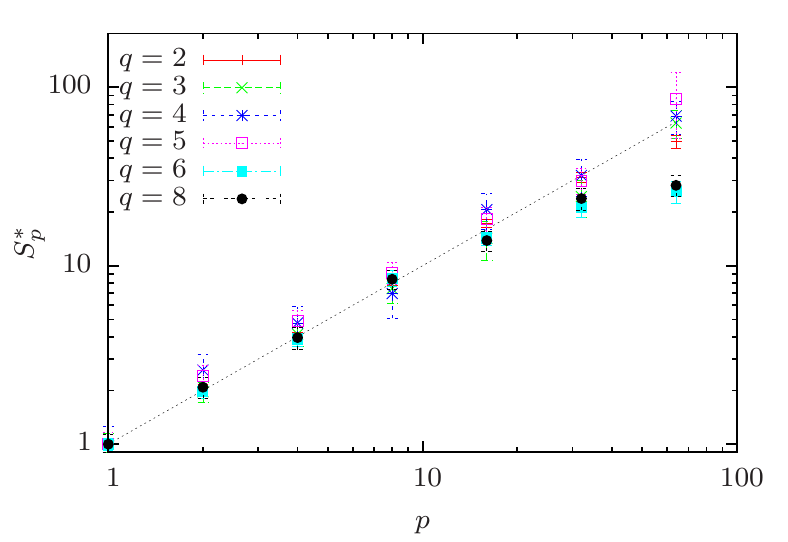}
  \caption{Statistical speedup for different $q$-state Potts models on a $64\times64$ square lattice. 
           For $q\le 4$ the Potts model exhibits a second-order phase transition, while 
           for $q>4$ the phase transition becomes first order.
          }
  \label{fig:qpotts}
\end{figure}

The $q$-state Potts model is furthermore well suited to take a look at the performance of the 
parallelization in the crossover regime from a second-order phase transition to a first-order 
phase transition. 
To this end, we considered extrapolated $M_{\rm opt}$ for different $q$ values. 
The result is shown in Fig.~\ref{fig:qpotts}. For $q\le 4$ the temperature-driven 
phase transition is of second-order and the scaling of the performance is very well. This still holds
for $q=5$ where a so-called weak first-order phase transition occurs. Already for $q=6$, we can see
the drop in performance for large degrees of parallelization, as before.

\begin{figure}
  \centering
  \includegraphics{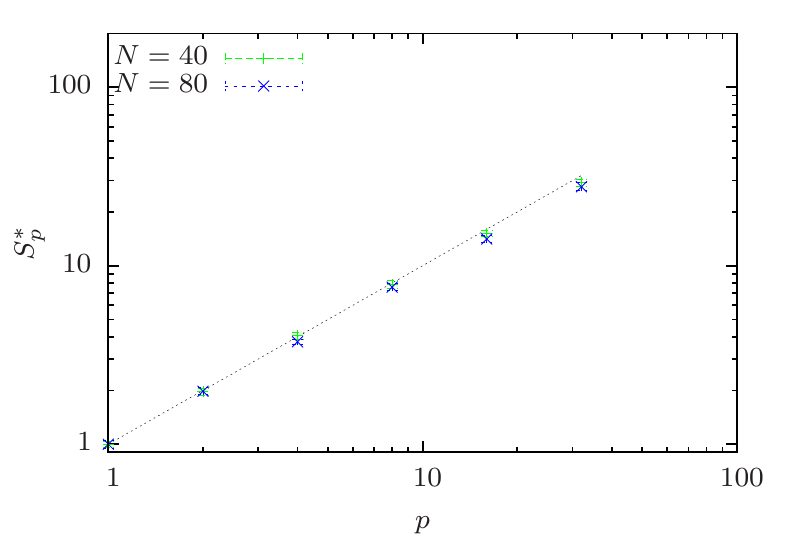}
  \caption{Statistical speedup for single bead-spring homopolymers of length $N=40,80$.
          }
  \label{fig:polymer}
\end{figure}

\vspace{1em}
Leaving the constraint of a lattice, we applied the parallel multicanonical method to a more complex
system, a single flexible bead-spring polymer. It consists of $N$ identical monomers, which are connected
to their bonded neighbors by a FENE spring potential and which interact with other monomers via a 
Lennard-Jones potential. The Hamiltonian is given by
\begin{equation}
  \mathcal{H} = 4\sum_{i=1}^{N-2}\sum_{j=i+2}^{N}\left( (\sigma/r_{ij})^{12}-(\sigma/r_{ij})^{6}\right)
                -\sum_{i=1}^{N-1}\frac{K}{2}R^2\ln\left(1-[(r_{i,i+1}-r_0)/R]^2\right),
\end{equation}
where $r_0$ is the average bond length, $\sigma=2^{-1/6}r_0$, $R^2=(r_{\rm max}-r_0)^2$, and $K$ is 
the spring constant. The parameters where chosen $K=40$, $r_0=0.7$, and $r_{\rm max}=1$ according 
to \cite{BeadSpringParameters}.
As mentioned above, the choice of updates is crucial. Here, we used a combination of single monomer
shift, spherical rotation, and double-bridging moves.
Using the example of a polymer system, we want to show the general applicability of the parallel
multicanonical method. Thus, we took an existing code of a multicanonical simulation with a fixed number
of sweeps per iteration and some thermalization between iterations.
The total number of sweeps per iteration was distributed onto the cores.
While we considered the full energy range for the spin systems, we restricted the multicanonical simulation
of the homo-polymer to an energy range around the collapse transition.
Figure \ref{fig:polymer} shows that the straightforward parallelization works also well for complex
off-lattice systems, which involve computationally more expensive energy calculations. Moreover,
this shows that the parallelization may be applied straightforwardly without taking too much care about
the involved parameters.

\vspace{1em}
In summary, the application of the parallel multicanonical method presented here is straightforward and
very efficient for a range of systems. We studied the performance on the example of the Ising model,
the $q$-state Potts model and a coarse-grained polymer model. Furthermore, we showed that the
parallelization may easily be adapted to flat histogram simulations in other ensembles. 
Overall, we could demonstrate a good performance yielding a close-to-perfect scaling $S^*_p\simeq p$
for up to $p=64$ cores.

\section*{Acknowledgments}
We are grateful for support from the SFB/TRR 102 (Project B04),
 the Leipzig Graduate School of Excellence GSC185 ``BuildMoNa'' and
 the Deutsch-Franz\"osische Hochschule (DFH-UFA) under grant No.\ CDFA-02-07.
J.Z. and M.M. were funded by the European Union and the Free State of Saxony.

\bibliographystyle{elsarticle-num}

\end{document}